# Modeling of 2D and 3D assemblies taking into account form errors of plane surfaces

**Serge Samper - Pierre-Antoine Adragna - Hugues Favreliere - Maurice Pillet**

*Université de Savoie – Polytech'Savoie – SYMME – B.P. 80439 – 74944 – Annecy le Vieux Cedex FRANCE*
*serge.samper@univ-savoie.fr; pierre-antoine.adragna@univ-savoie.fr; hugues.favreliere@univ-savoie.fr; maurice.pillet@univ-savoie.fr*

ABSTRACT.
*The tolerancing process links the virtual and the real worlds. From the former, tolerances define a variational geometrical language (geometric parameters). From the latter, there are values limiting those parameters. The beginning of a tolerancing process is in this duality. As high precisions assemblies cannot be analyzed with the assumption that form errors are negligible, we propose to apply this process to assemblies with form errors through a new way allowing to parameterize forms and solve their assemblies. The assembly process is calculated through a method allowing to solve the 3D assemblies of pairs of surfaces having form errors using a static equilibrium. We have built a geometrical model based on the modal shapes of the ideal surface. We compute the completely deterministic contact points between this pair of shapes according to a given assembly process. The solution gives an accurate evaluation of the assembly performance. Then we compare the results with or without taking into account form errors. When we analyze a batch of assemblies, the problem is to compute the Non-Conformity Rate (NCR) of a pilot production according to the functional requirements. We input probable errors of surfaces (position, orientation, and form) in our calculus and we evaluate the quality of the results compared to the functional requirements. The pilot production can then be validated or not.*

## 1 Introduction and purpose

Variational geometry is the first step of any tolerancing process even if most of the time this step is not conscious. Assumptions are made in order to simplify the "real" features. There are many ways to give geometry to some parameters. For example, we could say that a sphere could be given by one scalar if we do not care about its form errors. We could add to this parameter several errors like sphericity, cylindricity, conicity, and define deeply the geometry. In an extreme way, we could give the whole set of coordinates of the measured points as parameters. How many parameters do we need to define a given geometry? We propose a method allowing to define several levels to adjust the set of form parameters based on a geometric filtering.

A first answer is to start from the function from a top-down analysis (from the need to the solution). Then we could say that a minimum set of parameters could be sufficient, and the difference between the measured objects and the ideal has to be constrained within allowable values. We could want to know more about those deviations in order to better understand the influence of form errors on the function (finer analysis). The second answer is given by a bottom-up analysis where the manufacturer would like to know the forms of his production according to the customer's needs.

When the set of parameters is fixed (geometric model), the functional requirements (most of the time, precisions and assembly conditions) fix their limits. Tolerancing leads to quantifying the allowable limits fixed to those parameters. There are two ways to solve the compatibility of limits, the worst case and the statistical analyses. Simulations and industrial applications showed that statistical cases authorize larger tolerances. Thus, in this paper, we will show how we analyze a batch of assemblies by taking into account form errors. The aim is to validate a pilot production of assemblies.

In order to take into account form errors, the matting of surfaces has to be solved. If there are no defined criteria, the matting is undefined. We propose a method allowing to solve the matting of a pair of surfaces from the skin model by solving the static equilibrium of the pair depending on the matting force and the assembly device.

The solved values can come from a batch of measured surfaces of a pilot production that gives a batch of forms. In this paper, we use simulations that have the same statistic properties. Then we compute the mean and standard deviation of a probable batch of assemblies, which is checked with the functional requirements. This process gives NCR of a supplier assemblies according to the functional requirements of his customer taking into account full geometric deviations (dimension, position, orientation and form) of a pilot production in order to optimize the risks and costs of the function.

## 2 Assembly analysis model of form errors

### *2.1* *Literature review*

Parts assembly simulation is a heterogeneous research domain. Some papers are related to the stack of tolerances analysis but very few of them propose to take into account form errors.

Models for geometric description of features are the first step for a precision analysis. The complexity of the description is associated to the corresponding solving. From the most simple model based on a one-dimensional description assembly to a several dimensional one (discrete decomposition of features), we can find increasing data (and process) models. The one-dimensional models ([1],[2]) are used to solve size variations. With those scalar models the analyses can be done with the worst cases or the statistic solving.

In order to take into account orientations and positions of associated surfaces, literature on the subject gives several models using a set of parameters. Requicha [3] uses a geometric model with possible offsets of surfaces. Srinivasan [4] modified this model changing the offset by a sweeping sphere. Wirtz [5] introduced the vectorial parameter concept which Gaunet [6] added the TTRS [7] (Technologically and Topologically Related Surfaces) parameters of Clement. Sacks and Joskowicz [8] developed a method for mobile planar mechanisms. The Jacobian Torsor of Desrochers and Laperriere [9] is a kinematic model associated to an interval method. The T-MAP® (Tolerance Map) model of Davidson and Shah [10] allows to define those parameters in a constrain space. The Small Displacement Torsor (SDT) model of Bourdet and Clement [11] is a kinematic based model that unifies the three-translation and the three-rotation parameters. Giordano and Duret [12] [13] have developed the domain model in order to limit the SDT components to respect the zone concept of GPS standards [14].

The skin model [15] is a measurable surface, which must be parameterized (filtered) in order to take into account errors. The filtering can be made by several methods. Most of them use periodic functions that can build simple to complex shapes. Fourier series is the first used decomposition method used in order to define periods on circular ([16]) and linear shapes. Discrete harmonic transforms are well known in signal processing. 2D fourier is used by Capello et al. ([17]-[18]) used to define form parameters of rectangular shapes. Huang et al. [19] developed a mode-base method for form error decomposition and characterization by using Discrete Cosine Transform (DCT). In the case of disk shapes, the periodic method of Zernike Polynomials [20] can be very useful especially in optics.

A Tchebychef Fourier Series model is used by Henke ([21]) in order to describe the forms of a cylinder and identify specific types of error shapes. Some models based on fractals [22] and wavelets [23] give solutions on local simple geometry (roughness to waviness on rectangular shapes). In the case of cylindrical shapes, Fourier Tchebychef models used by Gouskov [24] defines specific form errors such as eccentric or conical.

We proposed in [25] a new way to define form error parameters based on the eigen-shapes of natural vibrations of surfaces. The originality is that the set of form parameters can be computed for any kind of shape. The modal shapes have vectorial space properties and are defined by using a discretization of the ideal surface. This method is resumed in section 2.3 .

In more general cases, some authors propose to define a basis built on the shape parameters of the CAD system as Cubeles ([26]) and Pottman ([27]) who propose to define a complex shape by a mechanism built on the set of control points of NURBS. Those methods based on polynomials can be used for a complex shape but they do not define systematically a basis with its properties and are difficult to use for a set of shapes. They are mainly used to define simple form error shapes.

Most of the models are top-down type ("a priori" shapes), they can be used to simulate shapes without knowing the production process. But some form error models uses bottom-up methods ("a posteriori" shapes) in order to use informations from measurements. A model of eigen-shapes derived from a Principal Component Analysis (PCA) is proposed by Summerhays and Henke in [21] and [28]. Camelio and Hu [29] use a PCA of a production data to study sheet metal assemblies. In order to analyze 3D medical images, Nastar, in [30] uses the modal shapes to define a set of instances (over time) of a human organ. In image processing [31] the PCA eigen-shapes is often used and gives interesting research directions.

In order to assess assemblies, some authors ([32],[33]) uses a form error zone which is added to a position and orientation zone. This type of method does not allow to mix position form and orientation and can increase costs. In order to analyze assemblies with a better accuracy, some works give matting solutions for circular [16] or prismatic joints [34]. The solution of matting between two surfaces is computed by minimization of a geometric criterion [16] [35]. Huang and Kong [36] use Monte Carlo method in order to define variations of assembled surfaces described by DCT. An other way is to study the static equilibrium of assemblies (with an assembly force) with form errors [37].

## 2.2 Position and orientation models: SDT domain model

The skin model is reduced to its associated surfaces [12]. A frame is attached to each surface. Their possible deviations are size, position and orientation. In order to model those last two types of deviations, we can use a kinematic based model (displacements parameters). Bourdet and Clement [11] defined the Small Displacement Torsor (SDT) that gives to positions and orientations of frames (of associated surfaces) a unified model. The SDT model allows calculating deviations without taking into account the observation frame. We can make simple operations on SDT (addition, transportation…).

The model of SDT domains of Giordano [12] is based on the duality of contact conditions and SDT components inequalities. It builds a mathematical representation of those inequalities by a convex hull in a space of 6 dimensions (or less as in the case of planar or axisymmetric mechanisms with 3D domains). Each specification or functional requirement is expressed as a domain. The functional requirement (FR) verifications consist in checking that the resulting SDT of all the involved deviations are within the FR domain (inclusion test in a 6D space).

## 2.3 Form errors parameters: modal model

### 2.3.1 Calculation of the modal model

We propose to define geometry variation by using a geometrical basis built on the mode shapes. They are calculated by using the dynamic equations of either continuum or discrete mechanics. We summarize those equations in the following.

$$\begin{cases} \rho \vec{\Gamma} = div([\sigma]) \text{ at each point of the volume} \\ [\sigma]\{n\} = \vec{f_n} \, dS \text{ at each point of the outer surface} \end{cases} \quad (1)$$

Equations (1) are the local dynamic equations and the associated boundary conditions, where $\rho$ is the density, $\vec{\Gamma}$ the acceleration vector, $\vec{f_v}$ is the field forces vector (nil here) $[\sigma]$ is the stress tensor, $\vec{n}$ the normal vector to the $dS$ local surface and $\vec{f_n}$ the applied local pressure on $dS$.

The solving of the set of equations (1) for a continuum domain leads to obtaining the displacement solutions. We can use this method for some simple geometry (as line, square or disks) cases [38]. In most of cases, there are no analytic solution. We have to use a numerical method in most cases. As we compute modal form bases for any type of shape, we use the Finite Element Analysis (FEA) [39] in order to determine the modal shapes.

The discrete analysis is resumed in the following equations (2-8).

The form error modes are the eigen-shapes of the perfect surface. If we use a FEA, the nodes of the meshing will be the corresponding measured points. The calculus of the natural mode shapes is made by the solving of the dynamic conservative equilibrium of the discrete geometry. The following equations (2) and the boundary conditions (which are described in displacements here) will give the eigen-shapes of the geometry.

$$M \ddot{q} + K q = 0 \quad (2)$$

$M$ is the mass matrix and $K$ the stiffness matrix of the structure and q the nodal displacements vector. The solutions of (2) are written as follows:

$$q_i = Q_i \, Cos(\omega_i t) \quad (3)$$

$Q_i$ is the amplitudes shape vector and $\omega_i$ the corresponding pulsation. The linear system (4) gives the $n$ natural pulsations $\omega_i$ and modes shapes $Q_i$.

$$\left( M^{-1} K - \frac{1}{\omega_i^2} I_d \right) Q_i = 0 \quad (4)$$

If there are no boundary conditions we obtain as rigid body modes as allowed global displacements. Those modes would be used, in our model, to define global position and orientation of the associated surface.

### 2.3.2 Projection operator

We suppose that we have measured (by a CMM for example) a given geometry in order to obtain a set of errors (displacements of measured points). This set of Euclidian displacement vectors becomes the *V* vector of the measured feature. If we do measure the position of nodes, we have to make an interpolation in order to set *V*. As we have defined the modal basis *Q*, we can obtain the projections of *V* in this basis by calculating the projection equation (5) in a non-orthonormal basis:

$$V = \sum_{i=1}^{n} \lambda_i Q_i = Q \lambda \tag{5}$$

Here, $Q$ is the matrix of eigen-vectors $Q_i$. $\lambda$ is the vector of modal coefficients $\lambda_i$. We choose to give $Q_i$ an infinity norm ($||Q_i||\infty = 1$) in order to give $\lambda_i$ a metric sense. We obtain the vector $\lambda$ by solving equations (6).

$$(Q^t Q)^{-1} Q^t V = \lambda \tag{6}$$

The $\lambda_i$ coefficient can be showed as a histogram chart called modal spectrum. This representation is understood for metrologists and designers.

We compute the vector residue of the projection in the reduced basis by

$$R(m) = V - \sum_{i=1}^{m} \lambda_i Q_i \quad \text{with } m<n \tag{7}$$

The corresponding scalar value is given by equation (8).

$$r(m) = ||R(m)|| \tag{8}$$

As the natural shapes are defined by modal masses and modal stiffness their complexity are globally growing with *i* (growing order of the frequency). This property is very useful in order to filter shapes, by keeping lowest values of $\lambda_i$. Most of measured shapes have got a decreasing modal spectrum.

### 2.3.3 Properties of modal bases

In Fig. 1 we show the first seventeen modes of a square plate with free boundary conditions. Solving can be made by a Finite Element Analysis (FEA) or with the help of analytical solutions [38]. The six first are rigid bodies that will reveal the position and orientation errors. We will convert them into SDT components of the associated surface. From the seventh, we can see form error modes. At a higher level of mode number, we will observe undulation and further, roughness. Form, undulation and roughness of a surface are distinguished by the level of periods of a flexure behavior. In addition, we could solve, size variations (depending on the size parameters) by using membrane behavior. We can see with this elementary example the properties of the modal bases. The mode shapes are sorted by the value of the associated natural frequency. As we do not care about this value, we keep this sorting operator that naturally gives mode shapes in increasing complexity order.

**Fig. 1 Mode shapes of a free BC's square**

In Fig. 1, a Finite Element Model (FEM) is shown, the square is discretized by 441 nodes (21*21). We could solve 2646 (441*6) mode shapes. Sorting them by increasing frequency give increasing complexity. Since the first six mode shapes are rigid body type, we can write a linear operator $[\alpha_{ij}]$ in order to represent the corresponding modal coefficients $\lambda_{Ri}$ into the equivalent SDT components (9). As there are 6 rigid body mode shapes for a free structure and 6 kinematic Degrees Of Freedom (DOF), $[\alpha_{ij}]$ is 6*6 constants. The $[\alpha_{ij}]$ matrix is non-singular and its opposite gives SDT by knowing the $\lambda_{Ri}$ coefficients. It is then easy to move the position and orientation of frames.

$$\begin{Bmatrix} \lambda_{R1} \\ \lambda_{R2} \\ \lambda_{R3} \\ \lambda_{R4} \\ \lambda_{R5} \\ \lambda_{R6} \end{Bmatrix} = \begin{bmatrix} \alpha_{11} & \alpha_{12} & \ldots & \alpha_{16} \\ \alpha_{21} & \alpha_{22} & \ldots & \alpha_{26} \\ . & . & \ldots & . \\ . & . & \ldots & . \\ . & . & \ldots & . \\ \alpha_{61} & \alpha_{62} & \ldots & \alpha_{66} \end{bmatrix} \begin{Bmatrix} Tx_A \\ Ty_A \\ Tz_A \\ Rx_A \\ Ry_A \\ Rz_A \end{Bmatrix} \quad (9)$$

The geometric parameterization built on mode shapes has the following properties.

**Geometric Basis:** $Q_i$ and $Q_j$ are independent if $i \neq j$
**Uniqueness:** To a set of coefficients, $\lambda_i$ corresponds only one form given by $V$
**Stability:** Continuum parameters (no conditional test, bifurcations, …).
**Growing complexity:** $Q_j$ is a more complex form than $Q_i$ if $j >> i$.
**Exhaustiveness:** Any variation of a shape can be decomposed in the modal basis
**Metric of shape variations:** $\lambda_i$ represents the metric coefficient of the mode shape $Q_i$ in the shape $V$

### 2.3.4 *Analytic or numeric method?*

We are making software that decomposes form errors of a measured shape. We can use analytic solutions of (1) in the cases of simple geometries. This way is easy to program, but in most of cases, we have to solve an FEM of eigen-shapes of a given geometry with the help of (4). Those shapes are input in a matrix $Q$ of vector parameters. This one can be an input of the software. Then the measured (or simulated) shape $V$ (thus numeric) decomposition is obtained by the same solving for the two types of eigen-shapes (if eigen-shapes are analytic we discretize them). If possible, we measure displacements of points that are taken from the FEM meshing (points are nodes). If not, we use interpolations.

### 2.3.5 *Separation of geometric errors*

Dimension, position and orientation errors are easy (as concepts) to separate to form, waviness and roughness. Form errors are not easy to define, compared to waviness and roughness. Those geometric errors can be defined by the same type of mathematic model but do not lead to the same mechanic behavior. Surfaces sizes can be very different compromising a size separation criterion. We propose to define form errors as the first mode shapes having the main influence on assembly results (i.e. final position of surfaces in contact). If the contact force is small and the surfaces are filtered to remove waviness and roughness, the two surfaces are assembled with a number of points equal to the possible components of the contact torsor. In the case of planes, we can apply one force and two moments thus there must be three contact points according to a small contact force and two form error surfaces.

The waviness is defined as intermediate mode shapes of the surface that are involved in a lot of small contact zones. They should remain smooth and could be deformed (elastically or plastically) with small forces.

Roughness is defined as numerous very small contact zones (spots) that should (in most of cases) have a very small influence on the assembly position.

In this paper we focus on position, orientation and form errors of surfaces.

### *2.4* *Single assembly of two form errors*

#### *2.4.1* *Contact determination*

First we make a filtering of each shape by keeping the firsts ($i<m$) coefficients $\lambda_i$. We obtain the deviation surface by subtracting the two matting shapes. We keep the m first coefficients considering the residue $r(m)$. We can then build a meshing adapted to $m$ (the Nyquist-Shannon evaluation is used in order to define a minimum number of FEM nodes). By this way, the number of possible matting facets is lower and their sizes are bigger.

Then we define a difference surface (surface #1 becomes perfect and #2 is the difference surface). In order to select all the possible matting facets, we compute the convex hull of this surface. The solving of the intersection of the matting force whit this convex hull gives the matting facet thus the contact configuration.

#### *2.4.2* *Position parameters and FR verification*

Assembly is obtained by knowing contact points and their displacements. The rigid coefficients $\lambda_{Ri}$ are deduced and equations (9) gives the corresponding SDT. As explained in 2.2 , we can verify by an inclusion test if those SDT components are inside the FR domain or not. The assembly of form errors can be compared with the assembly of associated surfaces (model without form errors) in order to know the influence of form errors on the functional requirement.

### *2.5* *Multiple assemblies verification and NCR*

Our proposal is to evaluate the assembly of two batches having form errors. The statistical characteristics of the batches are supposed to be known (from measurement of the pilot production batches for instance). Then, the assembly deviation or the NCR can be evaluated through simulations. The following part 2.5.2 presents a solution to simulate a batch of form error based on its statistical characteristics. As this paper deals with no measure, the part 2.5.1 presents our solution to simulate a virtual batch characteristics (mean and covariance matrix).

#### *2.5.1* *Generation of statistical characteristics of a virtual batch*

We have developed routines in order to make virtual skin models. We built virtual assemblies that have the properties of observed batches in order to explain our method even if we can study any kind of batch. The statistical description of a batch is based on a mean value of position, orientation and form and a covariance matrix of these parameters. The statistical characteristics can be obtained from modal descriptions of a measured or simulated batch. The modal characteristics of the simulated pilot production  (mean vector $\mu_\lambda$ and the covariance matrix $C_\lambda$ of modal coefficients) are obtained with the following method:
- A mean shape is identified, called the mother shape of the batch. Based on observed modal analyses, modal coefficients $\lambda_i$ follows a decreasing distribution law with a $\mu_0/i$ type evolution (in Fig. 2, $\mu_0$=0.2 mm). The number of modal coefficients is limited to *m* that could be associated to a measurement filtering.
- Based on this mother shape, a small-simulated batch is randomly drawn. The $\lambda_i$ modal coefficient are normally drawn with a $\sigma_0/i$ standard deviation (in Fig. 2, $\sigma_0$=0.01 mm). The size *n* of the simulated batch directly affects the covariance matrix. Because random modal coefficients are independently obtained, a batch of a large size of shapes will have a quite non-covariant covariance matrix. Then, the smaller *n* is ($n≥2$), the higher the covariance factors are.
- The mean modal signature $\mu_\lambda$ and covariance matrix $C_\lambda$ (Fig. 2) of the simulated batch are computed to create the simulated batch of *n* shapes.

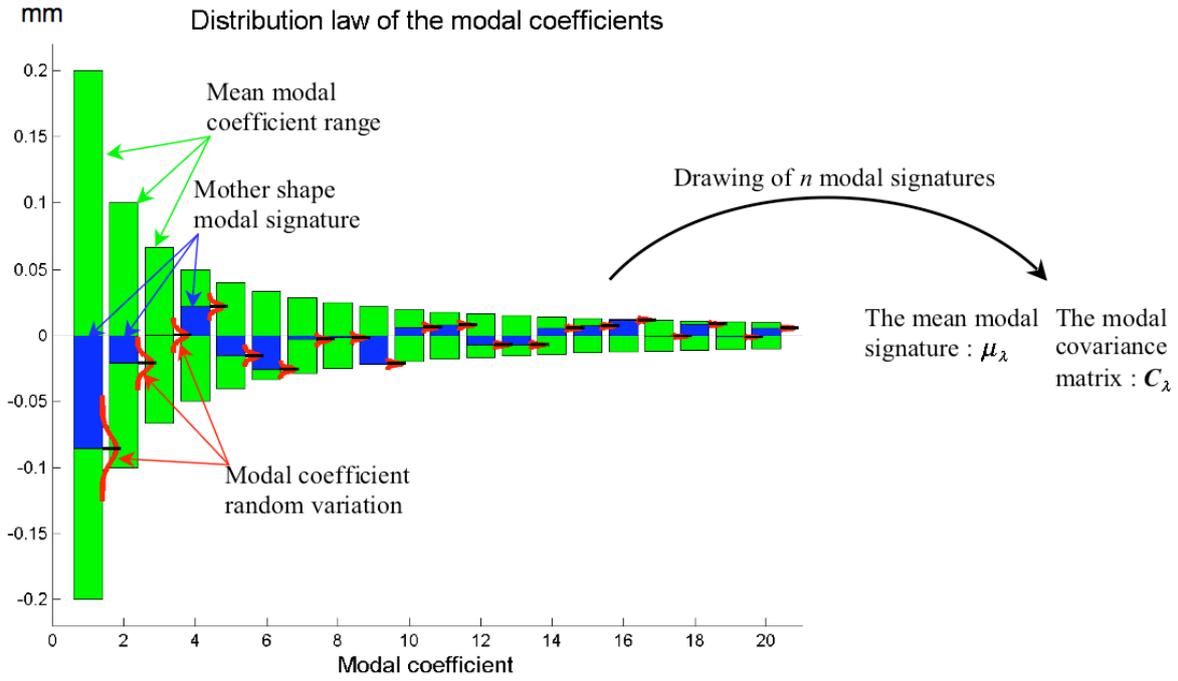

Fig. 2 Generation of the modal statistical characteristics of a virtual batch

### 2.5.2 Generation of a virtual batch based on statistical characteristics

As the aim is to check that the pilot productions respect the functional requirement, simulations are used. The NCR evaluation is based on the statistical characteristics of the pilot production. Our approach is to simulate two batches with a large number of parts based on the two statistical descriptions of the two pilot productions. One batch example is presented in Fig. 3.

The generation of one virtual batch is made as follows:
- A set of $N$ random modal signature $\Lambda_r$ (of $m$ modal coefficients) is drawn. Each modal coefficient follows a standard distribution law (for instance, normal distribution with a mean $\mu_{\lambda i} = 0$ and a standard deviation $\sigma_{\lambda i} = 1$).
- We compute a diagonal matrix $C_{\lambda \text{diag}}$ (that contains the variance on the main direction) and the transfer matrix $P$ (from the diagonal variance matrix to the batch covariance matrix) from the covariance matrix $C_\lambda$ of the pilot production batch. A modal signature $\Lambda_v$ of a virtual form error of the pilot production is obtained from the random modal signature $\Lambda_r$ with the following relation:

$$\Lambda_v = P \cdot \sqrt{C_{\lambda diag}} \cdot \Lambda_r + \mu_\lambda \quad (10)$$

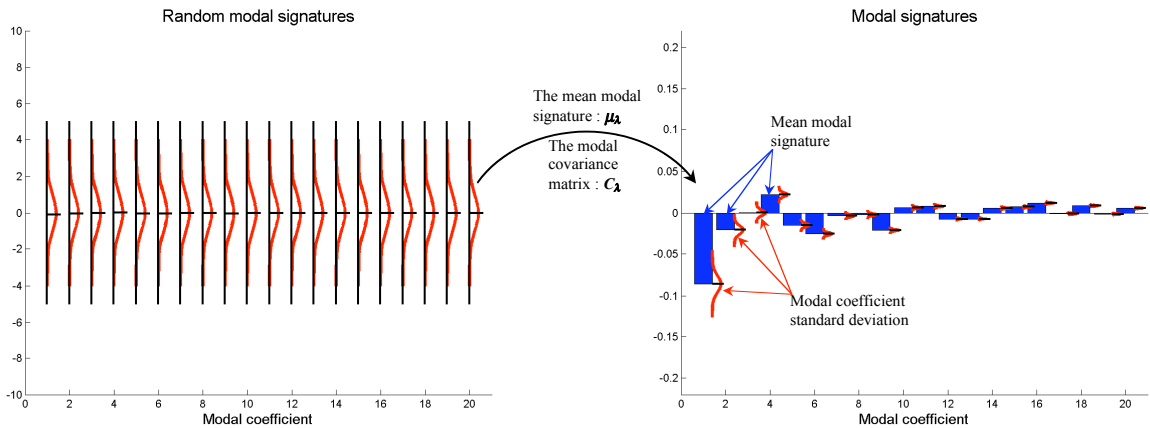

Fig. 3 Generation of a virtual batch based on statistical characteristics

The assemblies can be analyzed as presented in section 2.4. The NCR is computed by an inclusion test of the assembly SDT components inside of the FR domain.

## 3 Application on an assembly of two parts

The following example is a classic chain of parameters closed by a functional requirement ($FR_B$) of the $B$ pair of surfaces. The pair of surfaces $A$ ($A_1$ form part 1 and $A_2$ form part 2) presented in Fig. 4 has to keep a given distance. The location specification $t$ gives the value of the functional requirement (between surfaces $B$). We assume that the pair of faces of the functional requirement $B_i$ has no geometric errors. We will compute how deviations on the contact pair $A_i$ influence the relative positions of $B_i$. The positioning device is supposed to be perfect and the assembly is rigid.

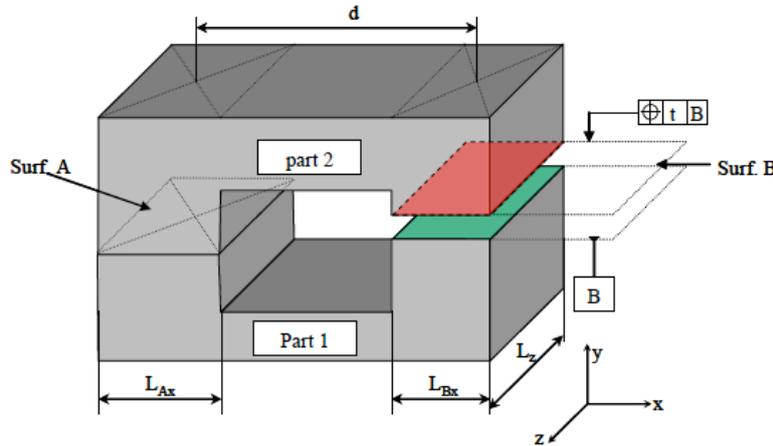

**Fig. 4 Functional Requirement of the assembly**

If the supplier is asked to give interchangeable parts 1 and 2, tolerancing for each part should be made and the corresponding NCR should be higher according to the same production than with the non-interchangeability one because some compatible form errors assemblies would not be allowed. In this paper, we focus on the taking into account of form errors on the assembly NCR without interchangeability condition.

In this example, the contact surface $A_i$ is a square which side has a length of 40 mm ($L_{AX}=L_z$). The functional requirement surfaces $B_i$ is a rectangle of 20*40 mm ($L_{BX} * L_z$) side lengths and the intermediate surface is a 40*40 mm square. The other dimensions of the parts do not influence our analysis.

The two parts are not supposed to be the same and have the same type of deviations. We propose to use the tolerances of Fig. 4.b. Then the question for the designer is "what minimum values can we input for the tolerance $t$ ?" The question for the producer is "are my products compatible with those specifications and those functions?". In order to help this relation between the customer and the supplier, we will make in the following some simulations of virtual assemblies.

We will analyze a simple assembly (Fig. 5) in order to show the method. The two matting surfaces $A_1$ and $A_2$, may have position, orientation and form errors. The matting position of a pair of parts is given by the position set composed of an external position device and a matting force. If there is no mobility of the assembly, this set stops six DOF. In the example shown here, the external device is composed of three rods that could give three contact points because the plane joint is dominating (3 fixed DOF). The position of the resultant force F determines the three other DOF. We use this principle in order to compute the relative position SDT of the parts according to the matting set and their form errors. We simulate probable form errors of $A_1$ and $A_2$ with decreasing values of their $\lambda_{1i}$ and $\lambda_{2i}$ coefficient according to their $i$ mode number (classic evolution of shapes).

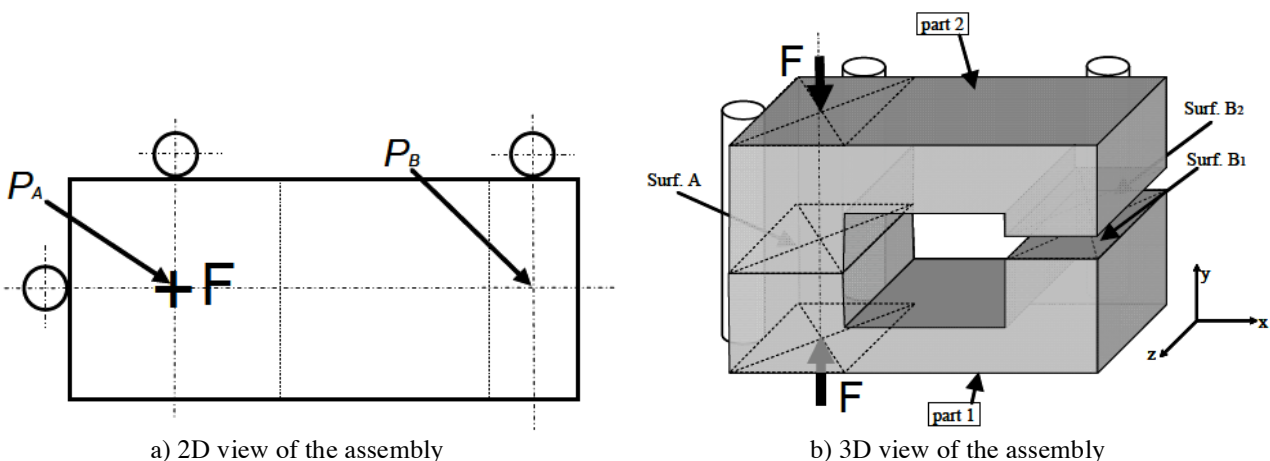

a) 2D view of the assembly        b) 3D view of the assembly

**Fig. 5 Assembly**

### 3.1 2D assembly

We assume here that geometric errors are nil in the *x* direction. In order to model position, orientation and form errors, we build the associated structural model by a free-free beam. As the measuring is on the *z* direction, the FEM DOF are *Ty* and *Rz* (translation about *y* and rotation around *z*). The six first modes are presented in fig. 6. The modes #1 and 2 are rigid in order to identify the position and rotation parameters. All the modes have an infinite norm; the maximum displacement of each one is equal to ±1. For example, mode #2 is a rigid body rotation and the maximum value of a node's displacement is 1. The modes represent possible forms of each surface.

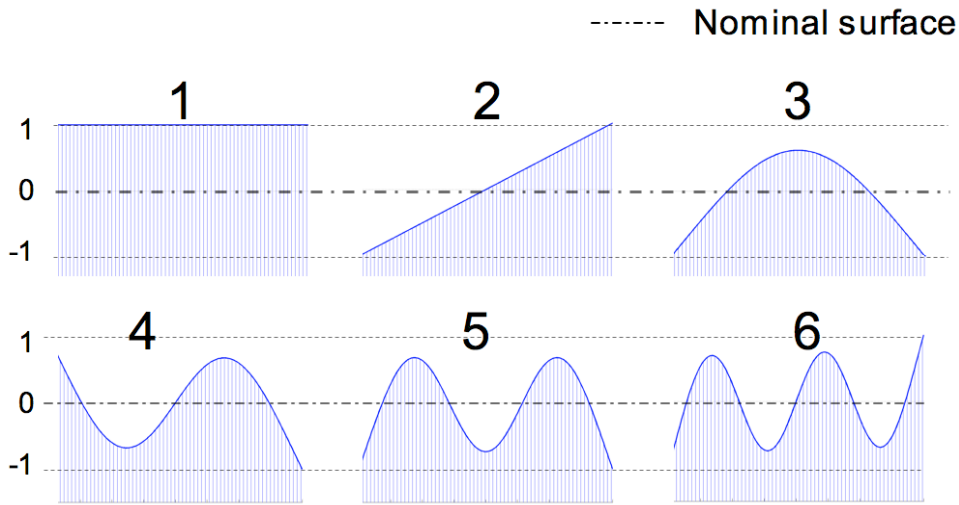

**Fig. 6 Mode-shapes of a free-free beam**

If we measure a form ("$A_1$" surface of Fig. 7) in order to extract its modal parameters, we obtain a vector of values shown here as a histogram. The most significant mode is the first mode (-0.01 mm amplitude of position), the second one is the third mode (-0.007 mm amplitude "banana" form), the third significant is the second (0.003 mm rotation amplitude) and the fourth significant is the ninth (0.003 mm amplitude). If we filter by using the lowest modes, we obtain the filtered shape of the focus in Fig. 7.b. This method uses the same scheme as Fourier and is well understood by geometric quality users.

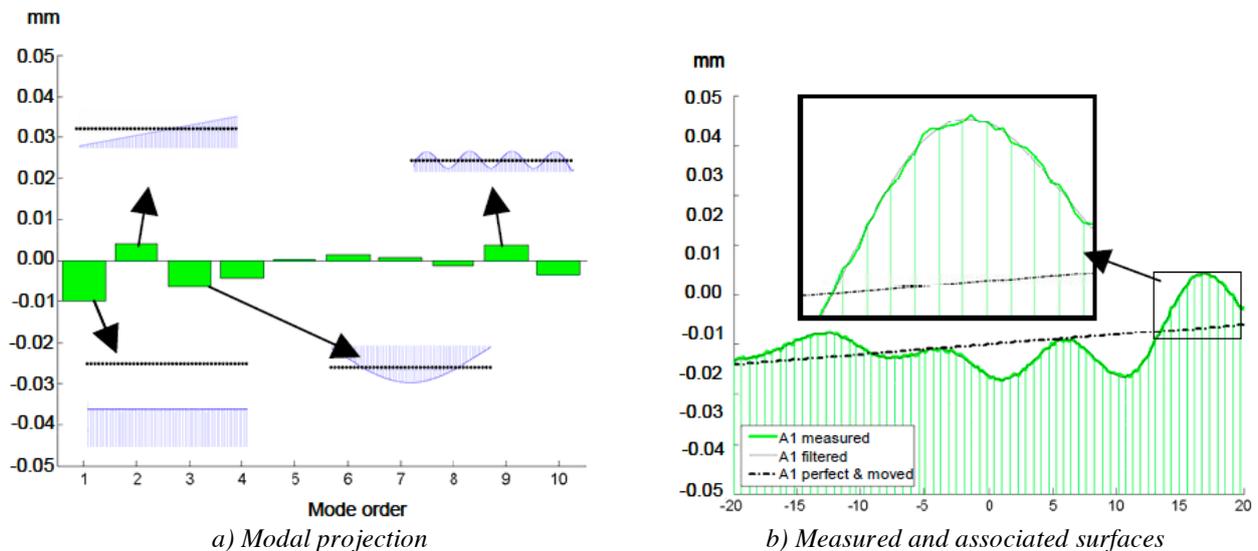

*a) Modal projection*  *b) Measured and associated surfaces*

**Fig. 7 Modal parameters**

#### 3.1.1 Contact facet

We assume that the contact between a pair of surfaces is given by a static solution. Thus we have to analyze the static equilibrium of the two parts linked by the matting surface. This one is ideally plane (in this example) and has three kinematics DOF in 3D analysis. The DOF are all removed by external links (fig. 5). A joint force *F* creates the matting pair.

Our approach consists on identifying all the possible contact points of the two faces regarding their form deviations and a given pre-positioning mechanism. A positioning force identifies the contact points that define the stable positioning. The first step of the method is the identification of the possible contact points. We define the difference surface $A$ linking surfaces $A_1$ and $A_2$. This surface corresponds to the difference of the corresponding form deviations. The set $\lambda_{A12}$ of modal parameters $\lambda_{1A2i}$ of the equivalent surface $A$ is obtained by the difference (eq. 11, Fig. 8) between the modal parameters $\lambda_{A2}$ (surface $A_2$) from $\lambda_{A1}$ (surface $A_1$).

$$\lambda_{1A2} = \lambda_{A2} - \lambda_{A1} \tag{11}$$

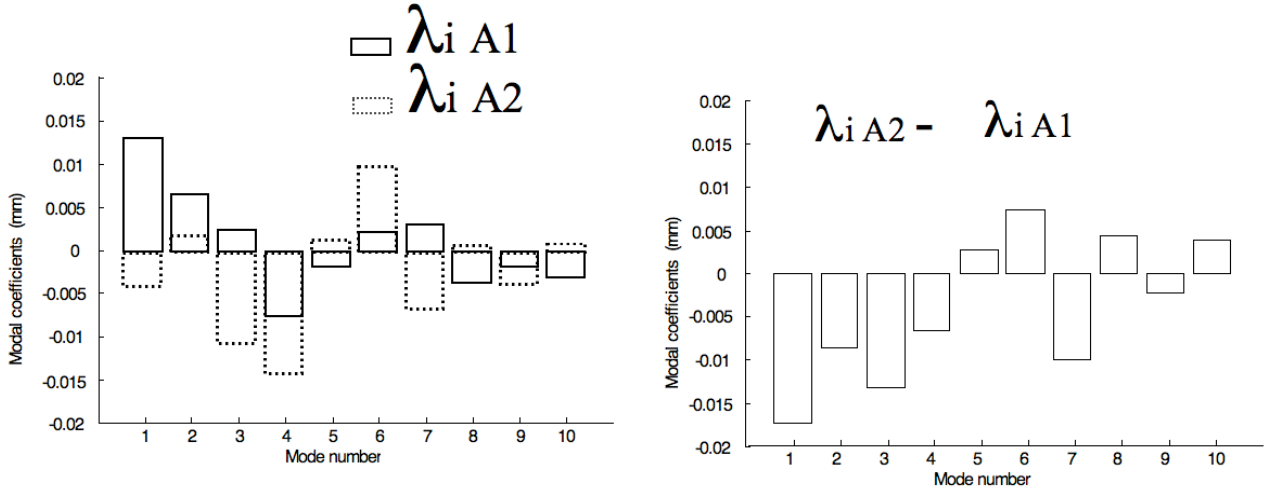

**Fig. 8 Modal coefficients of the difference surface**

On Fig. 9.a surfaces $A_1$ and $A_2$ in their measured frames are shown. Fig. 9.b presents the difference surface ($A_2$-$A_1$). By this way, we could consider that $A_1$ is perfect and $A_2$ has equivalent geometric errors. The second step of our approach is the identification of all (two here) the possible contact points between the two surfaces. The proposed solution is the use of the convex hull that identifies all the possible contact facets (segments here), hence all the possible contact points. Fig. 9.b shows the convex hull (polygon in 2D) of the difference surface (curve) and the potential contact points (small circles).

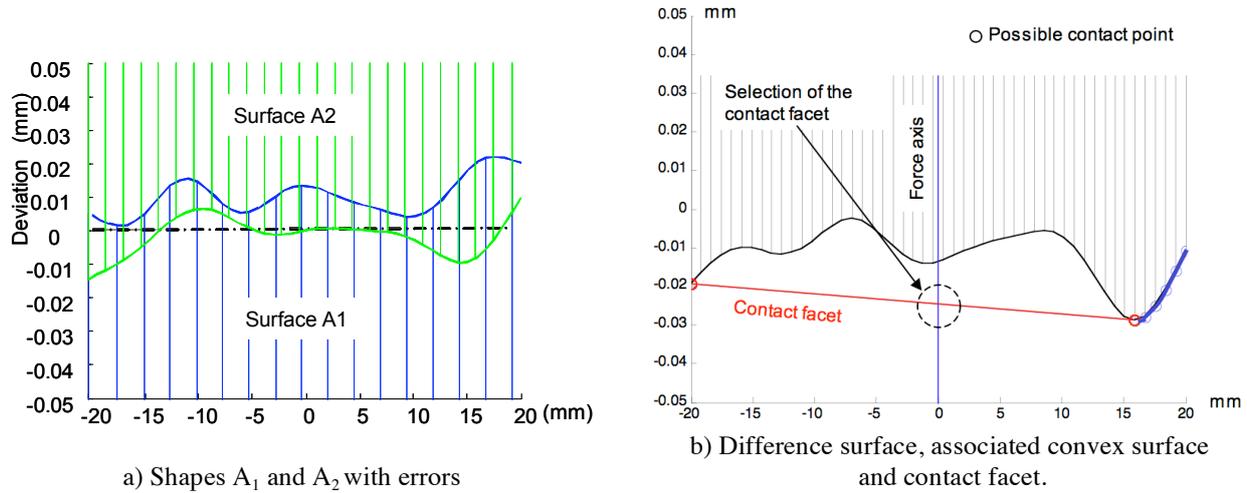

a) Shapes $A_1$ and $A_2$ with errors

b) Difference surface, associated convex surface and contact facet.

**Fig. 9 Contact facet determination**

When the convex surface is computed, the contact facet is selected by solving the intersection with the axis of the matting force.

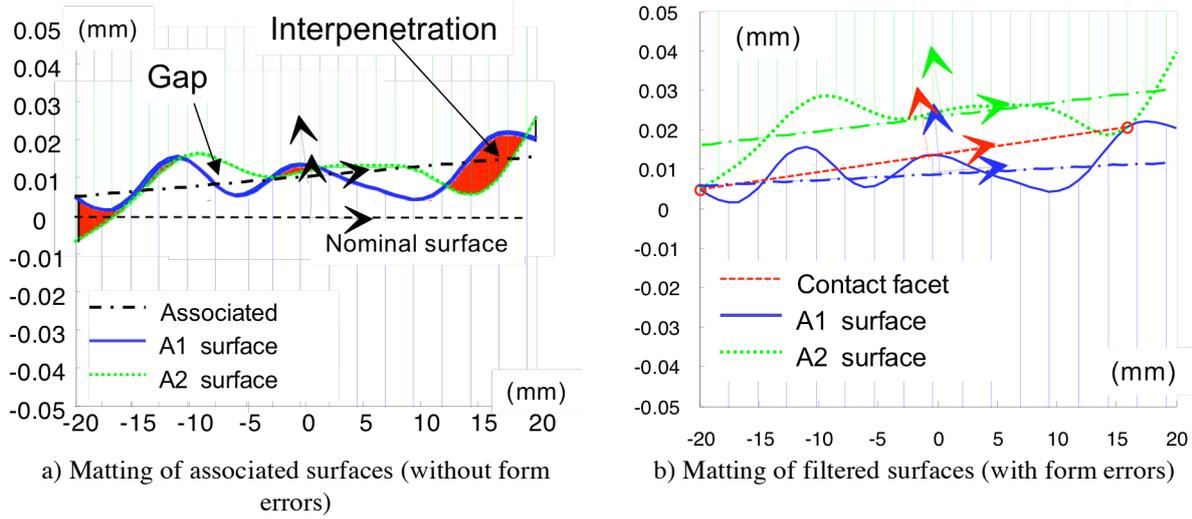

a) Matting of associated surfaces (without form errors)

b) Matting of filtered surfaces (with form errors)

**Fig. 10 Comparison between assembly methods**

In Fig. 10, we can observe assemblies with (Fig. 10.b) or without (Fig. 10.a) form errors. In Fig. 10.a we can see interpenetrations and gaps when associated surfaces are assembled (mean-square criterion). If we don't analyze form error contacts, we can either assemble mean-square surfaces or shifted associated surfaces, where the shift is two times the flatness value (one for each surface). Those two models enclose the values founded by our method.

*3.1.2  Form to position parameters*

By using equations (9), we translate the rigid $\lambda_i$ coefficients in the corresponding SDT. In Fig. 11 the SDT of associated surfaces $A_1$ ($E_{RA1}$) and $A_2$ ($E_{RA2}$) are relative deviations from the nominal surface. The SDT components of $E_{R1A2}$ (Error Rigid SDT) of $A_2$ from $A_1$ are shown in the 2D ($T_{yA}, R_{zA}$) space of SDT coordinates (Fig. 12.a). Here, we can observe the influence of the form error on the position of this plane joint.

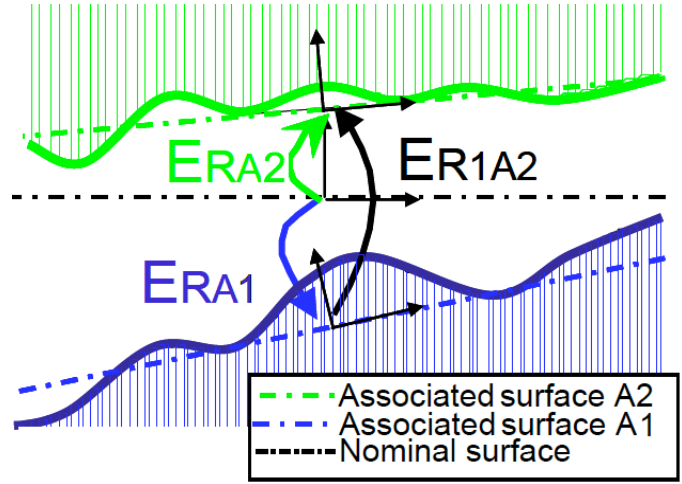

**Fig. 11 Deviations SDT of associated surfaces**

*3.1.3  Functional requirement verification of a single assembly*

In Fig. 12, we present the resulting SDT of the assembly of associated surfaces (without taking into account form errors) of $A_1$ and $A_2$. $E_{R1A2} = E_{RA1}-E_{RA2}$ is the rigid SDT of the assembly (without taking into account form errors). $E_{1A2}$ is the assembly SDT with form errors. We have added $E_{F1A2}$, the difference between $E_{1A2}$ and $E_{R1A2}$ in order to evaluate the taken into account form errors. As those torsors are calculated in the centre of surface $A$ and the functional $FR_B$ requirement is calculated in the centre of surface $B$, we move the $E_{1A2}$ SDT in order to be compared to the $FR_B$ domain. This domain is built by the four inequalities of each point of $B_2$ from $B_1$ (zone condition). If each (four) summit $S_{B2i}$ of $B_2$ perimeter must remain within the segment of $t$ range. The translation (along $y$ axis) of each summit (called $Ty_{SB2i}$) respects the following inequalities then $FR_B$ is verified.

$$-t/2 < Ty_{SB2i} < t/2 \qquad (12)$$

Those eight inequalities (12) are coupled so that four are sufficient. The corresponding $FR_B$ domain is shown Fig. 12.c. The two assembly SDT (with or without taking into account form errors) are first given in the local frame attached in the centre of the surface A (Fig. 12.a). The set of torsors $E_{X1A2}$ are transported in the same frame as the $FR_B$ domain (Fig. 12.b). The functional verification is made by an inclusion test showed in Fig. 12.c. In the presented case, deviations are compatible with the functional requirements.

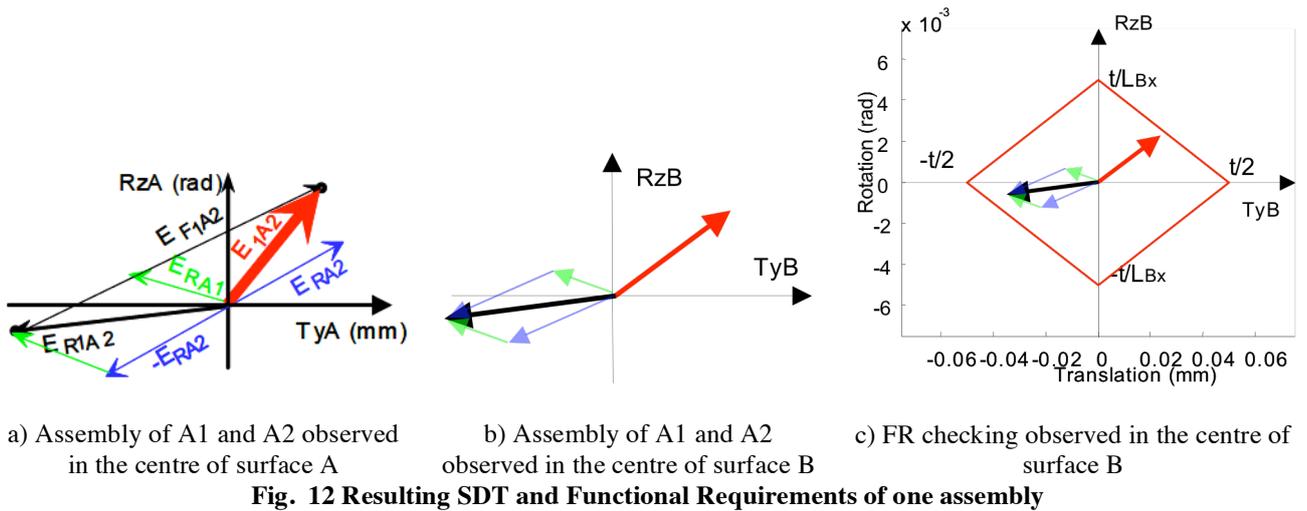

a) Assembly of A1 and A2 observed in the centre of surface A

b) Assembly of A1 and A2 observed in the centre of surface B

c) FR checking observed in the centre of surface B

**Fig. 12 Resulting SDT and Functional Requirements of one assembly**

The two SDT components verify the FR but we can see in this example that results are quite opposite.

*3.1.4  Functional requirements verification of a simulated pilot production*

The supplier has made a pilot production in order to check the NCR. He produces two independent batches of one hundred parts each (for part 1 and part 2). If we assume that any part 1 can be assembled with any part 2, we will have to test all the hundred assemblies.

We created simulated a set of random shape having decreasing values of modal coefficients $\lambda_i$ (an hyperbolic law was used) and covariances such as a batch production should have. This batch is usefull to assess our method but we can analyse other set of shapes. Cvetko *et al*. [40] gives an estimation of the NCR dispersion by the simulation dimension. In our example of 100 assemblies, the standard deviation of the NCR is about 0.5%.

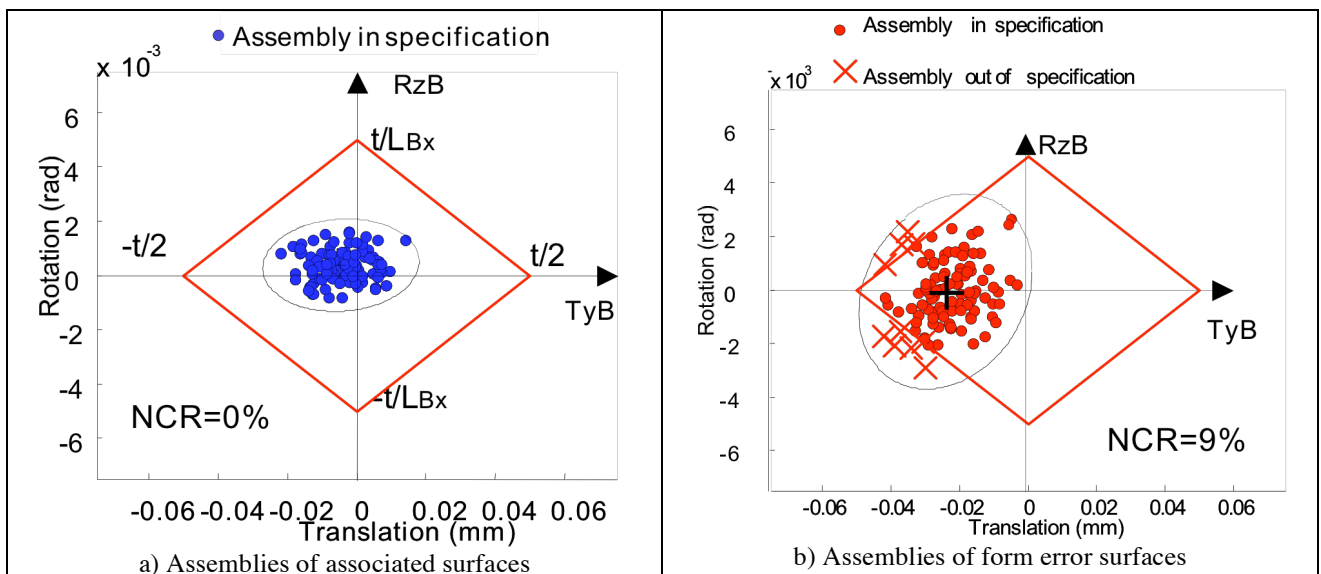

a) Assemblies of associated surfaces

b) Assemblies of form error surfaces

**Fig. 13 Conformity of a set of 100 assemblies**

Here, we fixed $t=0.1mm$ ($L=20mm$). As shown in Fig. 13, the conformity of this batch is better without taking into account form error in the assemblies. The risk is then to deliver some (9% here) non-conform assemblies without knowing the NCR. As those assemblies (Fig. 13.a) are made by using the associated surfaces (position, orientation but no form errors), the corresponding deviations are lower than those with form errors (Fig. 13.b). In addition, the results give the mean STD values (vertical cross in the centre of the ellipse) and the corresponding six-sigma ellipse. The supplier can identify the component involved in the NCR (here the mean of the batch is shifted on the translation direction at -0.025 mm). Here, the effective (with form errors) batch deviation could be inside the *FRB* if the mean were nil.

### 3.2 3D assemblies

The same assembly is presented with the assumption that the form error shape is a 3D one. The DOF are *Ty* (translation upon *Y* axis), *Rx* and *Rz* (rotations upon *X* and *Z* axis).

#### 3.2.1 Surfaces form errors

We have built one hundred different form errors for $A_1$ and $A_2$. The modal basis is computed by using the natural mode shapes of a free BC's (Fig. 1) of a square surface. We create probable (decreasing coefficients) form errors (Fig. 14.a,b,c) for $A_1$ and $A_2$. The three first $\lambda_i$ are associated to translations and rotations of surfaces. We solve the assembly of the associated surfaces (Fig. 14.d) in which form errors remains. Some interferences between $A_1$ and $A_2$ are shown in Fig. 14.d.

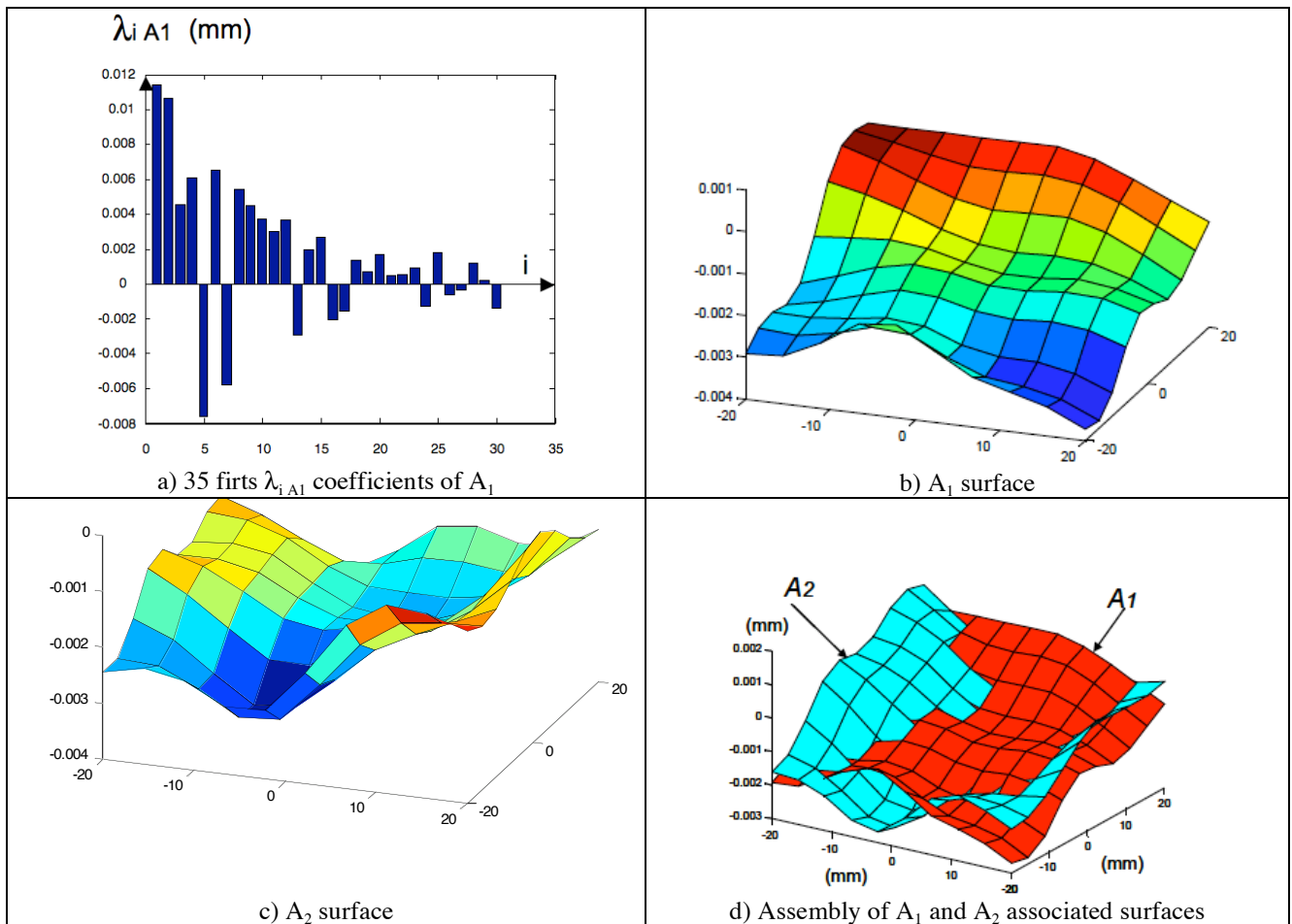

a) 35 firts $\lambda_{i\,A1}$ coefficients of $A_1$

b) $A_1$ surface

c) $A_2$ surface

d) Assembly of $A_1$ and $A_2$ associated surfaces

**Fig. 14 Surfaces A1, A2 and assembly of associated surfaces.**

#### 3.2.2 FCR verification of an assembly

We build a difference surface (Fig. 15.a) with the same method as the 2D one. A convex hull computation of this surface gives the set of possible contact facets. The intersection of the *F* axis with the convex hull gives the contact facet (thick triangle). Then we identify the three contact points belonging to this facet. Their *z* values give the three "rigid" coefficients of $A_2$ ($\lambda_{1\,A2}, \lambda_{2\,A2}, \lambda_{3\,A2}$). By moving $A_2$ from $A_1$ with those three coefficients, we obtain the matting conditions.

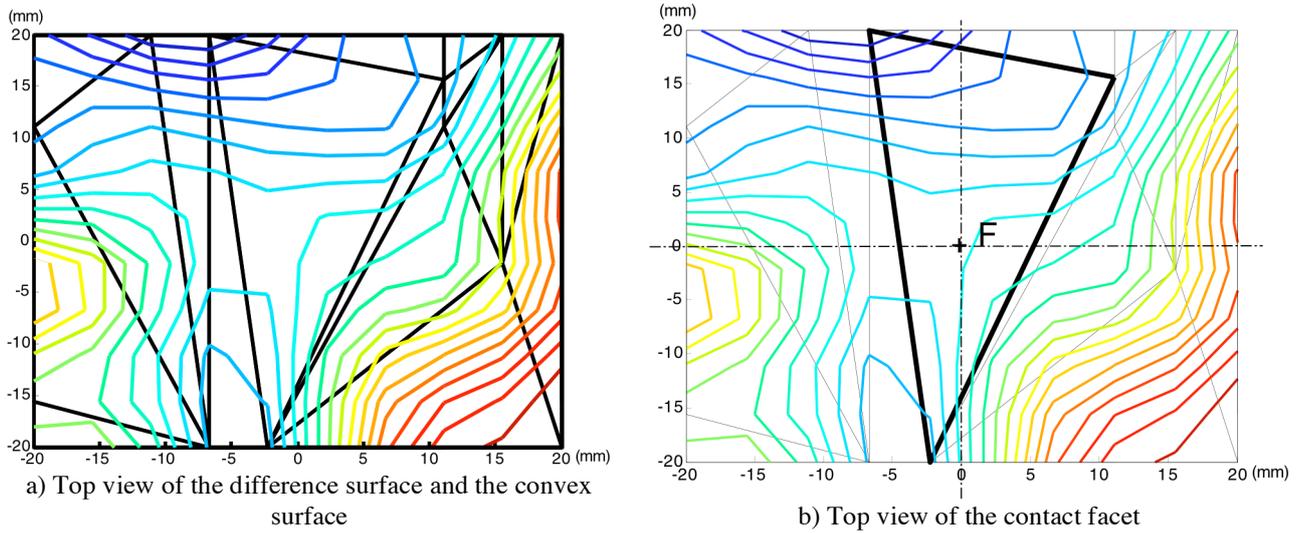

a) Top view of the difference surface and the convex surface

b) Top view of the contact facet

Fig. 15 Contact facet on the assembly

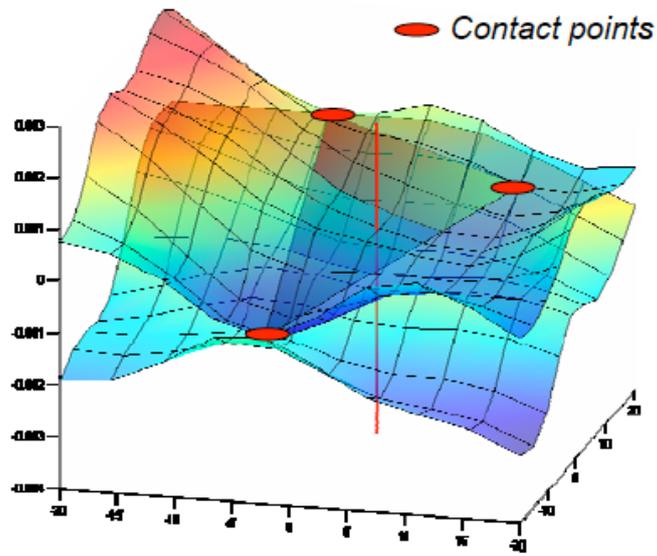

**Fig. 16 3D view of the contact**

In Fig. 15, the contact facet is defined by three points. In order to see both surfaces and the contact facet, we can see, in Fig. 17 the topographic top view of the two surfaces ($A_2$ in filled lines and $A_1$ in dashed lines) and the contact facet i.e. a triangle defined by the three summits ($P_1, P_2, P_3$). Surface $A_2$ is repositioned with the three ($\lambda_{1A2}, \lambda_{2A2}, \lambda_{3A2}$) coefficients that are solution of the contact solving. In the two zooms, we can observe that in $P_2$, the topographic lines have the same color (or gray) value (contact is effective) and in $P_4$, different values (gap).

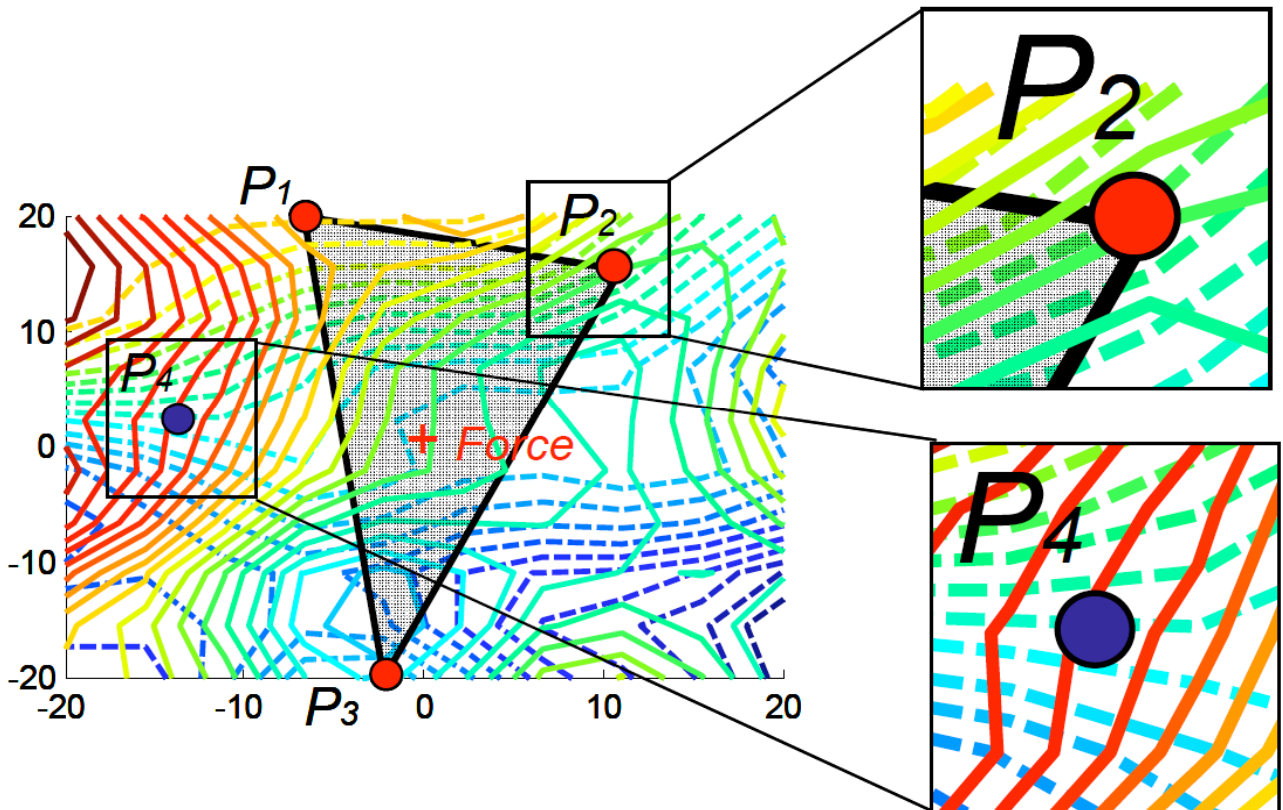

**Fig. 17 Assembled surfaces and contact points, a 2D view**

We calculate the SDT of relative position of $A_2$ from $A_1$ corresponding to those displacement values. Then we transport this torsor in the centre of the specification $B$. The functional requirements are respected if the SDT components are within the $FR_B$ domain (Fig 18). The assembly of form error surfaces is closer to the limits than the one with associated surfaces. It is easy to add SDT to several joints and check the $FR_B$.

The contact conditions of the $FR_B$ are given the by the eight inequalities of (12). Here, as there are three DOF parameters, those eight inequalities (half spaces) define an octahedron.

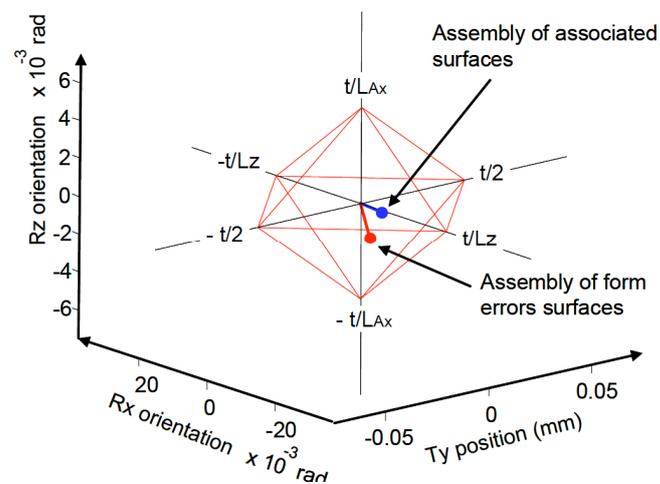

**Fig. 18 $FR_B$ domain and assembly verifications**

### 3.2.3 *Functional requirements verification of a pilot production*

After the virtual assemblies of one hundred pair of surfaces, we can compute the Non Conformity Rate of this batch. In Fig. 19 the two analyses of assemblies with or without taking into account form errors are shown (with mean SDT components and six sigma ellipsoid). In the case of associated surfaces, the whole batch is acceptable (within the $FR_B$ domain). The NCR of assemblies with form errors gives 14% out of $FR_B$ assemblies.

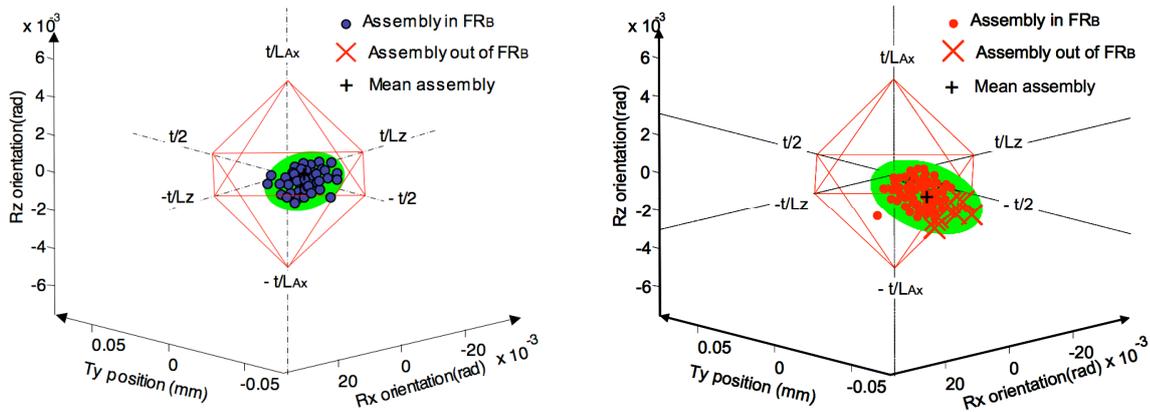

**Fig. 19 Functional requirement ($FC_B$) 3D Domain and SDT of assemblies**

**Remark:**
- In this article, we show several graphics in order to explain our method, but we do not need any of them for making our solving.
- As the modal description of form errors is easy to define for any shape (with the help of a FEA), this method can be extended to other matting assemblies. The FEA has to be made only one time for each type of geometry. The set of modal shapes is stored in a library of modal shapes used by form deviations software. We can then solve assemblies in order to obtain SDT of a set of assemblies and computing NCR according to FR's.

## 4 Conclusion

Assembly simulations are needed to analyze deviation influences on the functional requirements. We have presented how this goal can be reached taking into account form errors on a static mechanism. Surface form errors are defined by the modal method. The natural mode shapes of the theoretical feature automatically build a geometric basis with several interesting properties as automaticity, exhaustiveness and natural sorting of shape complexity. Any form error can be decomposed in this geometric basis. The modal coefficients are used as shape parameters of 2D or 3D features. Keeping the most important coefficients makes a filtering operation. In order to compute the assembly of two surfaces, we define a filtered equivalent form error. Contact points must be on their convex hull. The set of corresponding facets define all the possible contact configurations. Assembly determination is solved by a static method. The use of the force assembly axis intersecting the convex hull of the filtered equivalent form error selects the contact facet. Then SDT model is used to transport this contact deviation to the functional requirement surface. By the domain method, this result is compared to the functional requirement domain. A simulation of a pilot production shows Non Conformity Rate with or without taking into account form errors. The supplier can use this approach in order to better answer the customer needs (NCR mastering at a good cost). We are making routines in order to analyse several type of contact surfaces. Libraries of modal shapes are made in this aim.

**Acknowledgement**
The authors acknowledge the financial supports of the French Ministry of Research, the European Union, the "Assemblée des Pays de Savoie" and the Competitiveness Pole "Arve Industrie Haute Savoie Mont-Blanc".